\documentclass[twocolumn,preprintnumbers,amssymb,prb,superscriptaddress,floatfix]{revtex4}
\usepackage{graphicx}
\usepackage{dcolumn}
\usepackage{bm}
\usepackage{amsmath}
\usepackage{latexsym}
\usepackage[dvips]{color}
\usepackage{psfrag}
\begin{document}
\title{The path-coalescence transition and its applications}
\author{M. Wilkinson}
\affiliation{Faculty of Mathematics and Computing,
The Open University, Walton Hall, Milton Keynes, MK7 6AA, England}
\author{B. Mehlig}
\affiliation{Physics and Engineering Physics, Gothenburg University/Chalmers, 
Gothenburg, Sweden.}
\begin{abstract}
We analyse the motion of a system of particles subjected
a random force fluctuating in both space and time, and
experiencing viscous damping. 
When the damping exceeds a certain threshold, the system undergoes
a phase transition: the particle trajectories coalesce. 
We analyse this transition
by mapping it to a Kramers problem which we solve exactly.
In the limit of weak random force 
we characterise the dynamics 
by computing the rate at which caustics are crossed,
and the statistics of the particle density in the coalescing phase. 
Last but not least we describe possible
realisations of the effect, ranging from trajectories
of raindrops on glass surfaces to animal migration patterns.
\end{abstract}
\maketitle
\begin{figure}
\includegraphics[width=7.cm,clip]{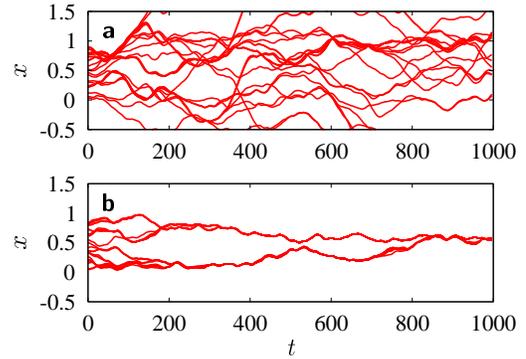}
\caption{\label{fig: 1}
Positions of $25$ particles with viscous dissipation
subjected to a spatially and temporally varying random force.
The dynamics are described by eqs.~(\ref{eq: 1}), (\ref{eq: 2}).
In both cases
$m=\tau=1$, $\xi=0.05$, $\epsilon=0.001$; {\bf a} $\gamma \approx 0.04$, 
{\bf b} $\gamma \approx 0.1$.}
\end{figure}

This paper discusses a surprising phase transition
illustrated in Fig.~\ref{fig: 1}. 
This figure shows
trajectories for a system
of particles with positions $x_i(t)$ subjected
to a random force fluctuating in both space and time,
and experiencing a resistive force proportional to
their velocities (the equations of motion
are defined by (\ref{eq: 1}) and (\ref{eq: 2}) below). The motion of any
one particle is obviously diffusive. Two particles with
very close positions and momenta must follow similar
trajectories, at least for a while. Diffusive motion
usually tends to reduce inhomogeneities in density,
and we might expect that the motion should resemble
the simulation in Fig.~\ref{fig: 1}a. However, a slight increase
in the resistive force causes a phase transition to the
situation shown in Fig.~\ref{fig: 1}b, which we term \lq path
coalescence'. This effect has been described in a paper
by Deutsch\cite{Deu85}, who
gave a theoretical analysis and numerical simulations indicating
the existence of a phase transition in this model.

Here we point out that these and related  equations of motion 
have a very broad range 
of applications in the physical
sciences, ranging from tracks of raindrops on randomly 
contaminated glass surfaces, to energetic electrons in 
disordered solids. In the \lq overdamped'
limit, where inertia is negligible, the model could describe
the response of animals to random fluctuations
of their environment. Possible applications of this type include
migration tracks of mammals and clustering of microorganisms.
Because of the large variety of applications we 
believe that the path-coalescence effect
deserves to be thoroughly understood. This paper
makes the following contributions.
First, adopting an approach \cite{hal65} used in
the theory of Anderson localisation,
we map the equations of motion to a Kramers problem
which we solve exactly. This enables us to  obtain
an exact criterion for the phase transition.
Second, 
we characterise
the dynamics by computing the rate at which trajectories (in
Fig.~\ref{fig: 1}) cross caustics. Third, in the coalescing
phase, we determine the statistics of the particle
density by calculating its pair-correlation function.
This allows us to deduce, at time $t$,
the expected number of particles condensing into a trail.
Fourth we argue that the model we consider
(which is more general than that put forward in \cite{Deu85})
exhibits a complex phase diagram. 
Finally we conclude with a discussion
of some of the applications of the path-coalescence effect.
Given the ubiquity of diffusive dynamics, the effect we 
describe here is bound to occur in a wide variety of different
contexts.

We consider trajectories for a system
of independent particles with positions $x_i(t)$, and momenta
$p_i(t)$. The equations of motion for any particle are
\begin{equation}
\label{eq: 1}
\dot x=p/m
\ ,\ \ \ 
\dot p=f(x,t)-\gamma p
\end{equation}
where $\dot x\!=\!{\rm d}x/{\rm d}t$, and $\gamma$ characterises
the strength of the viscous damping.
The statistical properties of the force $f(x,t)$ are translationally
invariant in both space and time, 
\begin{equation}
\label{eq: 2}
\langle f(x,t)\rangle\!=\!0\,,\,\,\,
\langle f(x,t)f(x',t')\rangle\!=\!c(x\!-\!x',t\!-\!t')\,.
\end{equation}
$\langle A\rangle$ is the ensemble average of $A$.
The correlation function decays rapidly as $\vert x-x'\vert \to \infty$
and as $\vert t-t'\vert \to \infty$. 
A suitable form (adopted in Fig.~\ref{fig: 1}) is
$c(\Delta x,\Delta t)\!=\!\epsilon^2 \exp(-\Delta x^2/2\xi^2)
\exp(-\Delta t^2/2\tau^2)$ where $\epsilon$ denotes the magnitude
of the force.
The dynamics of the model is characterised by 
two independent dimensionless variables:
$\chi\equiv\epsilon \tau^2/m\xi $ is a dimensionless measure of the strength of the random force, and 
the motion is overdamped when $\nu\equiv\gamma \tau \gg 1$.

{\sl Phase transition}. The phase transition 
is determined by the fraction of initial 
conditions for which the separation a pair of infinitesimally 
close trajectories approaches zero as $t\to \infty$: this 
is either $0$ or $1$. 
The separation $\delta x$ of two nearby trajectories has a lognormal
distribution. We have
\begin{equation}
\label{eq: 3}
\langle {\rm ln}|{\delta x(t)/{\delta x(0)}}|\rangle
=\lambda t\,,
\end{equation}
we call $\lambda$ the Liapunov exponent.
When $\lambda\!<\!0$, almost all nearby trajectories eventually
merge, conversely when $\lambda >0$ nearby trajectories 
almost certainly diverge. The condition for the phase
transition is therefore $\lambda=0$, and evaluation
of $\lambda $ is also valuable because it gives the rate
of coalescence.
 Linearising the equations of motion (\ref{eq: 1}) gives
 $\delta \dot x=\delta p/m$, and
 $\delta \dot p=-\gamma \delta p+\partial_x f(x,t)\,\delta x$.
%Linearising (\ref{eq: 1}) gives
%\begin{equation}
%\delta \dot x=\delta p/m\,,\quad\mbox{and}\quad
%\delta \dot p=-\gamma \delta p+\partial_x f(x,t)\,\delta x\,.
%\end{equation}
Here $\delta x$ and $\delta p$ are small separations between
pairs of trajectories.
As $t\to \infty$ the ratio between $\delta p$ and 
$\delta x$ has a stationary distribution.
We therefore write $X=\delta p/\delta x$, and find equations of motion
in terms of $X$ and $\delta x$:
\begin{eqnarray}
\label{eq: 5}
\delta \dot x&=&X\delta x/m\\
\label{eq: 6}
\dot X&=&\partial_x f(x(t),t)-\gamma X-X^2/m\,.
\end{eqnarray}
Since the distribution of $X$ remains stationary,
eq.~(\ref{eq: 5}) implies that
\begin{equation}
\label{eq:liap}
\lambda=\langle X \rangle/m\,.
\end{equation}

Consider the dynamics of $X$. When $-X$ is sufficiently
large, the noise term $\partial_xf\big(x(t),t\big)$ may be neglected and (\ref{eq: 6}) 
implies that $X$ reaches $-\infty$ in finite
time. This point is a caustic, where $\delta x$ passes through
zero and $X$ jumps from $-\infty$ to $+\infty$. 
As $t\to \infty$, one obtains a stationary
distribution, with $X$ going to $-\infty$
at a rate $-J$.

We now specialise to the case 
where the displacement of $X$ during the correlation time
$\tau $ is small compared to the deterministic terms ($\chi \ll 1$),
adopting the approach used in \cite{hal65}.
In this limit 
the probability density $P(X,t)$ for $X$ satisfies 
a Fokker-Planck equation \cite{Gar90}
\begin{equation}
\label{eq:fp}
\partial_t P=\partial_X[(X^2/m+\gamma X)P]+{\cal D}\,\partial^2_X P\,.
\end{equation}
with
diffusion constant
\begin{equation}
\label{eq: 8}
{\cal D}=-\frac{1}{2}\frac{\partial^2}{\partial x^2}
\int_{-\infty}^\infty {\rm d}t\ 
c(x,t)\bigg\vert_{x=0}
\ .
\end{equation}
We require a steady-state solution of eq.~(\ref{eq: 8}), $P(X)$, with a 
constant flux $J$, satisfying 
$ J=v(X)P-{\cal D}{{\rm d}P/{{\rm d}X}}$ with $v(X)=-{X^2/m}-\gamma X$.
It is convenient to express the solution in terms of dimensionless
variables: 
\begin{equation}
\label{eq: 10}
P(X)=\alpha\, \zeta(Z,\Omega)\ ,\ \ Z={X\over{(m{\cal D})^{1/3}}}\ 
,\ \ \Omega={\gamma m^{2/3}\over {{\cal D}^{1/3}}}
\end{equation}
where the constant $\alpha$ and function $\zeta$ are to be determined. 
The exact stationary solution of (\ref{eq:fp}) is
\begin{equation}
\label{eq: 11}
\zeta(Z,\Omega)=\exp\bigl[-\phi(Z,\Omega)\bigr]
\int_{-\infty}^Z{\rm d}Y\ 
\exp\bigl[\phi(Y,\Omega)\bigr]\,.
\end{equation}
Here $\phi(Z,\Omega)$ is proportional to the integral of $v(X)$
\begin{equation}
\label{eq: 12}
\phi(Z,\Omega)=-\frac{1}{\cal D}\int_0^X {\rm d}X'\ v(X')
={Z^3\over 3}+{\Omega Z^2\over 2}
\ .
\end{equation}
The flux $J$ is determined by normalisation.
\begin{figure}
\centerline{\includegraphics[width=7.cm,clip]{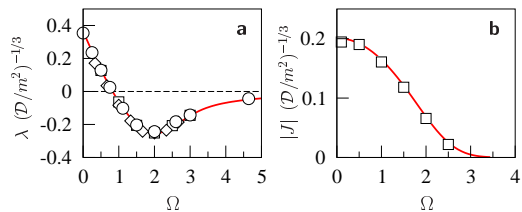}}
\caption{\label{fig:liap} 
{\bf a} Liapunov exponent: theory (line) and numerical experiments
for $\chi=0.1$ ($\diamond$), $0.02$ ($\Box$), and $0.001$ ($\circ$).
{\bf b} Rate $|J|$ of crossing of caustics,
theory (line), and numerical experiments for $\chi=0.02$ ($\Box$). }
\end{figure}
Using (\ref{eq:liap}) we obtain the Liapunov exponent 
\begin{equation}
\label{eq: 13}
\lambda=({\cal D}/m^2)^{1/3}\Psi_1(\Omega)/\Psi_0(\Omega)
\end{equation}
where
$\Psi_n(\Omega)=\int_{-\infty}^\infty {\rm d}Z\ Z^n\, \zeta(Z,\Omega)$.
The Liapunov exponent is shown in Fig.~\ref{fig:liap}a.
Consider the form of (\ref{eq: 11}). When $\Omega$ is large, the integral
is approximately constant over the interval between the maximum
of $\phi(Z,\Omega)$ at $Z=-\Omega$, and the point $Z={1\over 2}\Omega$
where $\phi$ is equal to its local maximum. In the interval
$[-\Omega,{1\over 2}\Omega]$, the function 
$\zeta(Z,\phi)\sim \beta \exp[-\phi(Z,\Omega]$ for some constant $\beta$.
Writing $\zeta(Z,\Omega)=\beta \exp(-\Omega Z^2/2)[1-Z^3/3+O(Z^4)]$, we
obtain 
\begin{equation}
\label{eq: 15}
\lambda=-\frac{1}{3} \biggl({{\cal D}\over m^2}\biggr)^{1/3}
\langle Z^4\rangle_0
=-{{\cal D}^{1/3}\over{m^{2/3}\Omega^2}}
=-{{\cal D}\over{\gamma^2 m^2}}
\end{equation}
where $\langle A\rangle_0$ defines the average of $A$ with the Gaussian
weight, $\zeta_0(Z)=\beta \exp(-\Omega Z^2/2)$.

Since $\lambda$ is negative for $\Omega\!\gg\! 1$, paths coalesce in this
regime. When $\Omega\!=\!0$, on the other hand, 
$\lambda=({\cal D}/m^2)^{1/3} \sqrt{3}\,\, 1\hspace*{-0.25mm}2^{5/6}\,
\Gamma(5/6)/(24\sqrt{\pi})>0$, which establishes
the existence of a non-coalescing phase for small $\Omega$
(at $\Omega\!=\!0$ the system is conservative; recent studies \cite{scho02}
of monodromy matrices for such systems obtain an expression
for $\lambda$ equivalent to ours at $\Omega=0$).
The transition between the two phases occurs at 
$\Omega_{\rm c}\approx 0.827$. 
The two examples shown in Fig.~\ref{fig: 1} are at
$\Omega=0.5$ and $\Omega=1.25$ respectively.
The damping which causes the maximum rate of coalescence
is determined by the minimum of 
$\Psi_1(\Omega)/\Psi_0(\Omega)$: this is at $\Omega \approx2.035$.

{\sl Rate of crossing of caustics}.
The magnitude of the flux determines
the rate at
which caustics appear on a given trajectory: 
$|J|\! =\!({\cal D}/m^2)^{1/3}/\Psi_0(\Omega)$, shown in
Fig.~\ref{fig:liap}b. $|J|$ is largest at $\Omega\!=\!0$,
$|J|\!=\!
({\cal D}/m^2)^{1/3}\,\Gamma(5/6)\,1\hspace*{-0.25mm}2^{5/6}/(8\pi^{3/2})$,
of the same order at $\Omega_{\rm c}$, and quickly drops to zero at large $\Omega$.
In Fig.~\ref{fig: 1}a caustics are still discernible.

{\sl Statistics of the particle density}.
In the overdamped limit ($\nu \gg 1$), the momentum is approximately
$p(t)=f(x(t),t)/\gamma$, so that eq.~(\ref{eq: 1}) 
approximated by
\begin{equation}
\label{eq: 16}
{{\rm d}x(t)\over{{\rm d}t}}
={1\over {m\gamma}}f(x(t),t)
\ .
\end{equation}
In this regime, a more concrete and complete understanding
of the path-coalescence effect is obtained
by considering
the statistics of the density of particles 
$\varrho(x,t)\!=\!\sum_i \delta \bigl(x_i(t)-x\bigr)$.
Translational invariance implies that an initially uniform 
density remains uniform,  $\langle \varrho(x,t)\rangle\!=\!\varrho_0$.
Path coalescence is revealed 
by the density-density correlation function 
${\cal K}(x,x';t)\!=\!\big\langle \varrho(x,t)\,\varrho(x',t)\big\rangle\!-\!
\varrho_0\delta(x-x')$.
Because of translational invariance,
the correlation function ${\cal K}$ is a function of $\Delta x\!=\!x\!-\!x'$ 
only: 
we write ${\cal K}(x,x';t)\!=\!K(\Delta x,t)$. 
A tendency for particles to cluster is demonstrated
by ${K}(\Delta x,t)$ becoming large for $\Delta x$ small, in the 
limit $t\to \infty$. 

When $\chi \ll 1$, 
we find that the correlation function satisfies a 
Fokker-Planck equation\cite{Gar90}
\begin{equation}
\label{eq: 17}
{\partial K(\Delta x,t)\over{\partial t}}
={\partial^2\over{\partial \Delta x^2}}
\bigl[D(\Delta x)\,K(\Delta x,t)\bigr]
\ .
\end{equation}
The diffusion constant
\begin{equation}
\label{eq: 18}
D(\Delta x)={1\over{m^2\gamma^2}}\int_{-\infty}^\infty {\rm d}t\ 
[c(0,t)-c(\Delta x,t)]
\end{equation}
approaches zero quadratically
at the origin: $D(\Delta x)\sim \kappa \Delta x^2$ for $\Delta x\ll \xi$. 
When $\Delta x \gg \xi$ the diffusion constant approaches a constant value 
$D_0$. 

Now consider the properties of solutions of 
eq.~(\ref{eq: 17}). We note that eq.~(\ref{eq: 17})
is in the form of a continuity equation, 
$\partial K/\partial t+\partial j/\partial \Delta x=0$, so that the
integral of the correlation function over all $\Delta x$ is a conserved
quantity. The flux of the correlation function passing the
separation parameter $\Delta x$ at time $t$ is
\begin{equation}
\label{eq: 19}
j(\Delta x,t)=-{\partial \over{\partial \Delta x}}
\big[D(\Delta x)K(\Delta x,t)\big]
\ .
\end{equation}
Consider an initially uniform distribution of density, with
value $\varrho_0$
(corresponding to $K(\Delta x,0)=\varrho_0^2$). For $\Delta x \ll \xi$ 
the diffusion constant is an increasing function of $\Delta x$. 
Together with (\ref{eq: 19}) this implies an initial flux of 
correlation towards $\Delta x=0$. At large times, 
$K(\Delta x,t)$ is thus sharply peaked at the origin. 
For $\Delta x\gg \xi$, on the other hand, the approximate solution of 
(\ref{eq: 17}) is
\begin{equation}
\label{eq: 20}
K(\Delta x,t)\sim \varrho_0^2\,
{\rm erf}\biggl({\vert \Delta x\vert \over{\sqrt{4D_0 t}}}\biggr)
\ .
\end{equation}
Using the fact that $K(\Delta x,t)$ satisfies a conservation
law, we deduce that the average number of particles condensing
into each trail at time $t$ is
\begin{equation}
\label{eq: 21}
N(t)\sim {4\over{\sqrt{\pi}}}\varrho_0\sqrt{D_0 t}
\ .
\end{equation}
When $t/\tau$ is large, and
$\Delta x\ll \xi$ (but not too close to zero)
the flux $j(\Delta ,t)$ is found to be approximately uniform. This implies
\begin{equation}
\label{eq: 22}
K(\Delta x,t)\sim {\varrho_0^2\over{\kappa}}\sqrt{D_0\over{\pi}}
{1\over{\Delta x\sqrt t}}
\ .
\end{equation}
The $1/X$-divergence of (\ref{eq: 22})
is non-integrable, so that this expression must fail 
near the origin.
An exact calculation shows that $K(0,t)=\exp(2\kappa t)$. 

\begin{figure}
\includegraphics[width=7cm,clip]{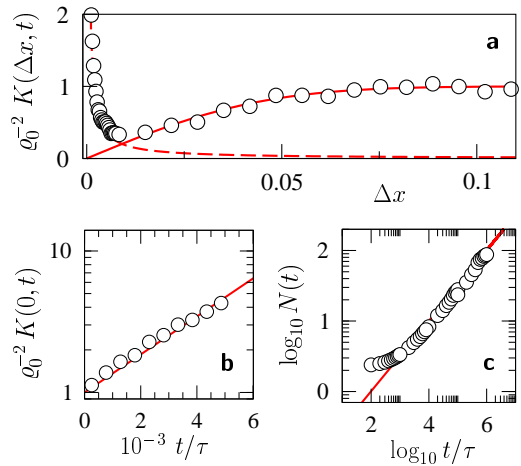}
\caption{\label{fig: 2}
Statistics of $\varrho(x,t)$ in the coalescing
phase.
{\bf a} Density correlation $K(\Delta x,t)$ of the process
(\ref{eq: 23}): numerical experiments    
$(\circ)$, limiting theoretical forms (\ref{eq: 20}), full line, 
and (\ref{eq: 22}), dashed line. 
Parameter values: $\varepsilon^2 
\approx 1.25\times10^{-8}$, $\xi \approx 6.4\times10^{-3}$, 
and $t=5\times 10^5\tau$.  {\bf b} Numerical results verifying
that $K(0,t)=\exp(2\kappa t)$ for the same parameter values.
 {\bf c} Mean cluster size $N(t)$,  of the process (\ref{eq: 23}):
 numerical experiments  ($\circ$), theory (\ref{eq: 21}), line.
Parameter values: $\varepsilon^2 \approx 2\times10^{-7}$, 
$\xi \approx 1.6\times10^{-3}$. Particles are considered
to be part of a cluster of $N$ if their positions
are within an interval of length $\xi$.}
\end{figure}

{\sl Tracking-minima regime}.
When $\chi\gg 1$, there is an overdamped phase in which the particles
follow minima of the random potential obtained by integrating
the force with respect to $x$. A particle executes a rapid jump
to a lower minimum whenever the minimum that it has been tracking 
disappears.

{\sl Discrete model.} The simplest model exhibiting    
the path-coalescence effect is a discrete-time
random walk which approximates (\ref{eq: 16}):
\begin{equation}
\label{eq: 23}
x_i(t+\tau)=x_i(t)+F_n(x_i(t))
\end{equation}
with $t=n\tau$. By analogy with (\ref{eq: 2}) we take 
$\langle F_n(x)\rangle=0$
and $\langle F_n(x)F_{n'}(x')\rangle=\delta_{nn'}C(x-x')$.
Deutsch\cite{deutsch2} gave a detailed analysis of (\ref{eq: 23}).
Here we remark that (\ref{eq: 23}) allows us
to obtain the simplest possible understanding of why coalescence occurs.  
Consider the linearisation of (\ref{eq: 23}):
\begin{equation}
\label{eq: 24}
\lambda = t^{-1}\,\langle\log | \delta x(t) |/|\delta
x(0)|\rangle
={\langle \log \vert 1+F'_n\vert \rangle /\tau}\,.
\end{equation}
When     
the magnitudes of the derivatives
$F'_n ={\rm d}F_n/{\rm d}x$ are small compared to unity, Taylor expansion 
of the logarithm gives 
$\lambda \sim -{1\over 2}\langle {F_n^\prime}^2\rangle/\tau$, so that 
 $\lambda $ is negative. This approximation is equivalent to (\ref{eq: 15}).
This argument indicates that the coalescence is a second-order effect.
Also, if the random displacements are larger than their
correlation length, the coalescence effect disappears: 
if $F_n$ has a Gaussian distribution, (\ref{eq: 24}) shows that
$\lambda$ becomes positive when 
$\langle {F_n^\prime}^2\rangle$ exceeds $2.421\ldots$.
Fig.~\ref{fig: 2} shows numerical results confirming 
eqs.~(\ref{eq: 20}), (\ref{eq: 21}) and (\ref{eq: 22}),
using numerical simulations of the mapping (\ref{eq: 23}).

{\sl Applications.} In the remainder we discuss 
a number of possible realisations of the path-coalescence effect. 

One example is the motion of liquid droplets on a surface, 
moving in one direction under a constant force (rain blown off a perspex windshield is an example
of this situation). If the surface is randomly contaminated, the wetting
angle will be different on opposite sides of each drop, and the trajectory
of the droplet will be randomly deflected. We assume 
the surface contaminants are smeared over an area
large compared to that of the droplets 
(perhaps resulting from cleaning the windshield with a waxy polish), 
so that nearby droplets are deflected
in the same direction. There is no interaction between the drops unless
they are close enough to combine due to surface tension: we stress
that the coalescence is that of the paths taken by different drops,
not of the droplets themselves. We model the motion of a droplet
across the surface by a particle of mass $m$.
At position ${\bf r}=(x,y)$ on the surface, the drop is 
subject to a force ${\bf F}({\bf r})+F_0\, {\bf j}$, where $F_0$ is the 
magnitude of a steady force acting in the direction of the unit 
vector ${\bf j}$ defining the $y$-axis and ${\bf F}({\bf r})$ 
is a homogeneous and isotropic random force with correlation length $\xi$. 
We assume that the particles are subjected to a viscous 
resistive force proportional to their velocity across the surface, 
such that the equations of motion are
\begin{equation}
\label{eq: 25}
m{{\rm d}{\bf r}\over{{\rm d}t}}={\bf p}\ ,\ \ \ 
{{\rm d}{\bf p}\over{{\rm d}t}}
=F_0 \,{\bf j}+{\bf F}({\bf r})-\gamma {\bf p}
\end{equation}
(where ${\bf p}$ is the momentum of the drop).
When the fluctuating force is weak, the 
trajectories are locally approximated by straight lines, 
with $x$ approximately constant and 
with $y$ increasing at a rate $v_y=F_0/m\gamma$. 
The equation of motion in the direction transverse
to the constant force is then in the form of equation
(\ref{eq: 1}), with $f(x,t)=F_x(x,v_yt)$.
In the case where the motion of the droplets is
sufficiently damped, the trajectories coalesce
onto fixed trails.

The path-coalescence effect may also be relevant to the motion
of energetic electrons in disordered solids. For example,
the effect may be relevant to a \lq branching'
observed in the flow of electrons away from a constriction
in a two-dimensional electron gas with very low scattering \cite{Top01}.
The experiment shows regions of markedly increased current
density persisting to some distance from the constriction.
This was explained by showing similarity
to simulations of independent electron 
motion in the smoothly varying random potential
of the doping atoms. This model is essentially (\ref{eq: 25}), 
with $\gamma=0$. 
Theoretical discussion of this system \cite{Kap02} has emphasised that 
caustics are important in understanding the 
empirical results. 
We remark that these experiments might show an even more 
pronounced effect if dissipation
were introduced. Fig.~\ref{fig: 1}a is similar to the 
flows discussed in \cite{Top01}, showing the branching effect 
and caustics. Adding dissipation to
the electron motion would cause the paths of the
electrons to coalesce, as in Fig.~\ref{fig: 1}b.
We note that 
dissipation of the electron motion could be increased by
increasing the temperature of the system. As well as giving 
a criterion for the phase transition, our expression for $J$ also
gives a quantitative prediction for the rate of formation
of caustics along a given trajectory.

There are also potential applications of the overdamped equations, 
(\ref{eq: 16}) or (\ref{eq: 23}), in the biological
sciences, involving the movement of organisms in response
to small random fluctuations in their environment. The 
model provides a mechanism through which large
numbers of organisms can congregate without communicating.
One example of this type is the migration
of animals across a nearly homogeneous smooth terrain. 
Thus Fig.~\ref{fig: 1}b could be thought of as a map showing paths
of animals on an Eastward migration.
The paths of the animals will be deflected by small
random fluctuations of topography or vegetation: 
the animals can be drawn 
together onto the same paths,
even if there is no communication between them, and
no gross features in the terrain favouring particular
routes. 
A second example applies to simple   
organisms such as plankton which can move in response to
changes in their environment, such as nutrient concentration. 
In cases where there
are small, spatially correlated random fluctuations of the 
nutrient concentration, the path-coalescence effect could lead
to unexpectedly large concentrations of organisms. Such
a mechanism could have been utilised by evolution, enabling 
simple organisms which cannot communicate directly 
to congregate for sexual reproduction.

\noindent {\sl Acknowledgement}.  Anders Eriksson
(Gothenburg University) 
suggested to us that the path-coalescence effect might be 
relevant to animal migration patterns. Financial support
from Vetenskapsr\aa{}det is gratefully acknowledged.


\begin{thebibliography}{4}
\bibitem{Deu85}
J. Deutsch, {J. Phys. A: Math. Gen.} {\bf 18}, 1449 (1985);
Phys. Rev. Lett. {\bf 52}, 1230 (1984).
\bibitem{hal65} B. I. Halperin, Phys. Rev. {\bf 139}, A104 (1965).
\bibitem{Gar90}
C. W. Gardiner, {\sl Handbook of Stochastic Methods, 2nd ed.},
Spinger, New York (1990).
\bibitem{scho02} H. Schomerus and M. Titov, Phys. Rev. E {\bf 66},
066207 (2002), and references cited therein.
\bibitem{deutsch2} J. Deutsch, {J. Phys. A: Math. Gen.} {\bf 18}, 1457
(1985).
\bibitem{Top01}
M. A. Topinka, {\sl et al.} {Nature} {\bf 410}, 183-6, (2001).
\bibitem{Kap02}
L. Kaplan, {Phys. Rev. Lett.} {\bf 89}, 184103 (2002).
\end{thebibliography}
\end{document}